\documentclass[twoside,twocolumn,9pt]{article}
\usepackage{extsizes}
\usepackage[super,sort&compress,comma]{natbib} 
\usepackage[version=3]{mhchem}
\usepackage[left=1.5cm, right=1.5cm, top=1.785cm, bottom=2.0cm]{geometry}
\usepackage{balance}
\usepackage{times,mathptmx}
\usepackage{sectsty}
\usepackage{graphicx} 
\usepackage{lastpage}
\usepackage[format=plain,justification=justified,singlelinecheck=false,font={stretch=1.125,small,sf},labelfont=bf,labelsep=space]{caption}
\usepackage{float}
\usepackage{fancyhdr}
\usepackage{fnpos}
\usepackage[english]{babel}
\addto{\captionsenglish}{%
}
\usepackage{array}
\usepackage{droidsans}
\usepackage{charter}
\usepackage[T1]{fontenc}
\usepackage[usenames,dvipsnames]{xcolor}
\usepackage{setspace}
\usepackage[compact]{titlesec}
\usepackage{hyperref}
\usepackage{bm}
\usepackage{amssymb}

\usepackage{epstopdf}%This line makes .eps figures into .pdf - please comment out if not required.

\begin{document}

%\raggedbottom

\pagestyle{fancy}
\thispagestyle{plain}
\fancypagestyle{plain}{

%%%HEADER%%%
\renewcommand{\headrulewidth}{0pt}
}
%%%END OF HEADER%%%

%%%PAGE SETUP - Please do not change any commands within this section%%%
\makeFNbottom
\makeatletter
\renewcommand\LARGE{\@setfontsize\LARGE{15pt}{17}}
\renewcommand\Large{\@setfontsize\Large{12pt}{14}}
\renewcommand\large{\@setfontsize\large{10pt}{12}}
\renewcommand\footnotesize{\@setfontsize\footnotesize{7pt}{10}}
\makeatother

\renewcommand{\thefootnote}{\fnsymbol{footnote}}
\renewcommand\footnoterule{\vspace*{1pt}% 
\color{gray}\hrule width 3.5in height 0.4pt \color{black}\vspace*{5pt}} 
\setcounter{secnumdepth}{5}

\makeatletter 
\renewcommand\@biblabel[1]{#1}            
\renewcommand\@makefntext[1]% 
{\noindent\makebox[0pt][r]{\@thefnmark\,}#1}
\makeatother 
\renewcommand{\figurename}{\small{Fig.}~}
\sectionfont{\sffamily\Large}
\subsectionfont{\normalsize}
\subsubsectionfont{\bf}
\setstretch{1.125} %In particular, please do not alter this line.
\setlength{\skip\footins}{0.8cm}
\setlength{\footnotesep}{0.25cm}
\setlength{\jot}{10pt}
\titlespacing*{\section}{0pt}{4pt}{4pt}
\titlespacing*{\subsection}{0pt}{15pt}{1pt}
%%%END OF PAGE SETUP%%%

%%%FOOTER%%%
\fancyfoot{}
\fancyfoot[RO]{\footnotesize{\thepage}}
\fancyfoot[LE]{\footnotesize{\thepage}}
\fancyhead{}
\renewcommand{\headrulewidth}{0pt} 
\renewcommand{\footrulewidth}{0pt}
\setlength{\arrayrulewidth}{1pt}
\setlength{\columnsep}{6.5mm}
\setlength\bibsep{1pt}
%%%END OF FOOTER%%%

%%%FIGURE SETUP - please do not change any commands within this section%%%
\makeatletter 
\newlength{\figrulesep} 
\setlength{\figrulesep}{0.5\textfloatsep} 
\makeatother
%%%END OF FIGURE SETUP%%%

%%%MY COMMANDS%%%
\newcommand{\Wi}{\mathrm{Wi}}
\newcommand{\Wicr}{\mathrm{Wi}_{\mathrm{cr}}}
\newcommand{\pst}{P_{\mathrm{st}}}

%%%TITLE, AUTHORS AND ABSTRACT%%%
\twocolumn[
  \begin{@twocolumnfalse}
\vspace{3cm}
\sffamily
\begin{tabular}{m{4.5cm} p{13.5cm} }

& \noindent\LARGE{\textbf{Effect of internal friction on the coil--stretch transition in turbulent flows}} \\%Article title goes here instead of the text "This is the title"
\vspace{0.3cm} & \vspace{0.3cm} \\

 & \noindent\large{Dario Vincenzi\textit{$^{a,\ddag}$}}\\ % Full Name,\textit{$^{b}$} and Full Name\textit{$^{a}$}} \\%Author names go here instead of "Full name", etc.

 & \noindent\normalsize{\it Universit\'e C\^ote d'Azur, CNRS, LJAD, Nice, France}

\\

 & \noindent\normalsize{A polymer in a turbulent flow undergoes the coil--stretch transition when the Weissenberg number,
\textit{i.e.}~the product of the Lyapunov exponent of the flow and the relaxation time of the polymer, surpasses a critical value. 
The effect of internal friction on the transition is studied by means of Brownian dynamics simulations of the elastic dumbbell model in a 
homogeneous and isotropic, incompressible, turbulent flow and analytical calculations for a stochastic velocity gradient.
The results are explained by adapting the large deviations theory of Balkovsky \textit{et al.} [\textit{Phys. Rev. Lett.}, 2000, \textbf{84}, 4765]
to an elastic dumbbell with internal viscosity.
In turbulent flows, a distinctive feature of the probability distribution of polymer extensions is its power-law behaviour for 
extensions greater than the equilibrium length and smaller than the contour length.
It is shown that although
internal friction does not modify the critical Weissenberg number for the coil--stretch transition, it
makes the slope of the probability distribution steeper, thus
rendering the transition sharper. Internal friction therefore provides a 
possible explanation for the steepness of the distribution of 
polymer extensions observed in experiments at large Weissenberg numbers.
} \\%The abstrast goes here instead of the text "The abstract should be..."

\end{tabular}

 \end{@twocolumnfalse} \vspace{0.6cm}

  ]
%%%END OF TITLE, AUTHORS AND ABSTRACT%%%

%%%FONT SETUP - please do not change any commands within this section
\renewcommand*\rmdefault{bch}\normalfont\upshape
\rmfamily
\section*{}
\vspace{-1cm}

%%%FOOTNOTES%%%

\footnotetext{\textit{$^{a}$~E-mail: dario.vincenzi@univ-cotedazur.fr}}
%\footnotetext{\textit{$^{b}$~Address, Address, Town, Country. }}

%Please use \dag to cite the ESI in the main text of the article.
%If you article does not have ESI please remove the the \dag symbol from the title and the footnotetext below.
%\footnotetext{\dag~Electronic Supplementary Information (ESI) available: [details of any supplementary information available should be included here]. See DOI: 00.0000/00000000.}
%additional addresses can be cited as above using the lower-case letters, c, d, e... If all authors are from the same address, no letter is required

\footnotetext{\ddag~Associate, International Centre for Theoretical Sciences, Tata Institute of Fundamental Research, Bangalore, India}

%\ddag~Additional footnotes to the title and authors can be included \textit{e.g.}\ `Present address:' or `These authors contributed equally to this work' as above using the symbols: \ddag, \textsection, and \P. Please place the appropriate symbol next to the author's name and include a \texttt{\textbackslash footnotetext} entry in the the correct place in the list.}

%%%END OF FOOTNOTES%%%

%%%MAIN TEXT%%%%

\section{Introduction}
\label{sect:introduction}

\thispagestyle{empty} %TO BE REMOVED

The coil--stretch transition is the complete unravelling
%a strong increase of the end-to-end length 
of a polymer that occurs when the polymer
is immersed in a non-uniform flow field and the magnitude of the velocity gradient surpasses
a critical value. It was initially predicted\cite{deGennes,h75} and observed experimentally\cite{psc97,sbsc03}
in a laminar extensional flow. 
The essential features of the coil--stretch transition, such as the strong distortion of the polymer
and the associated conformational hysteresis, can be predicted\cite{deGennes,h75} by using
a model as simple as the elastic dumbbell, which
consists of two inertialess beads connected by a spring. Moreover, 
if the contour length of the polymer is used as fitting parameter,
the dumbbell model satisfactorily reproduces the experimental measurements of the 
end-to-end distance.\cite{psc97,lpsc97,lhssc99,ssc04} 
References~\citenum{nk99,l05,s05,s18} contain a comprehensive
review of single-polymer dynamics in laminar flows.

It was later discovered\cite{bfl00,eks02,gcs05} that the 
coil--stretch transition also occurs in chaotic or turbulent flows, albeit with partially different features.
The most notable difference between extensional and turbulent flows is in the probability distribution of the polymer end-to-end distances.
%In turbulent flows, indeed, the core of the distribution behaves as a power of the end-to-end length.
In turbulent flows, indeed, the core of the distribution displays a power-law behaviour,
which indicates that a wide range of polymer extensions is observed even when the magnitude of the velocity gradient is very large.
This feature of the statistics of the end-to-end distance
 was predicted by applying large-deviations techniques to the dumbbel model in a random flow\cite{bfl00} and was observed
in both microfluidics experiments\cite{gcs05,ls10,ls14} and
numerical simulations of turbulent flows\cite{eks02,bcm03,pt05,wg10,bmpb12,gpp15} (see also Ref.~\citenum{bc18}, for a review).

The dynamics of a polymer involves internal dissipation processes, generally referred to as `internal friction',
which originate from local energy barriers
to short-range conformational changes, such as bond rotations, or from interactions between distant segments of the polymer that
come close in space.\cite{deGennes79}
%Although its microscopic origin is not fully understood,
%internal friction is known to play an important role in several phenomena, such as protein folding, ...
In coarse-grained models of elastic polymers, such as the bead--spring chain,\cite{bird,o96} 
internal friction
has been introduced by adding a linear `dashpot' to each elastic link, which yields a resistive
force proportional to the rate of deformation of the link. This idea was 
proposed by Kuhn and Kuhn~\cite{kuhn} under the name of `internal viscosity'. 
The early applications 
of internal viscosity were mainly concerned with
the rheology of viscoelastic fluids (see Refs.~\citenum{mw85,bird,l88,bo92,ry06} and references therein).
For instance, internal viscosity is known to cause shear thinning.\cite{l88}
More recently, there has been renewed interest
in bead--spring models with internal viscosity thanks to their application to the study of biopolymer dynamics
(see, \textit{e.g.}, Refs.~\citenum{ptw01,kMcl07,ssbdn12,chm13,sc16}).

The effect of internal viscosity on polymer stretching has been studied for
laminar, planar velocity fields.\cite{fl81,mw89,s93,w93} %\citenum{bw74,fl81,mcaw86,mw89,s93,w93}.
In particular, it was shown in Ref.~\citenum{fl81} that a moderate internal viscosity reduces 
the steady-state end-to-end distance, although without affecting
the critical velocity gradient for the coil--stretch transition. In contrast, when the magnitude of internal viscosity
exceeds a threshold value, polymers hardly deform. The limiting case of a purely extensional
flow was shown to be special, since in such a flow
internal viscosity does not modify the steady-state configuration of the polymer.
The goal of this study is to examine the effect of internal friction
on the coil--stretch transition when the velocity field is turbulent.

It ought to be noted that the notion of internal viscosity has been subject to some criticism,\cite{fl81,ds98,l05} for 
the magnitude of the force exerted by the dashpot
is not easily estimated from the molecular properties of the polymer and
it has been difficult %to identify measurable macroscopic phenomena that could be unambiguously attributed to internal viscosity alone
to find conclusive experimental evidence for the need of internal viscosity in bead--spring chains
(the reader is referred to Ref.~\citenum{kcp18} for a recent introduction on the notion of internal friction and the
use of internal viscosity in polymer models). Nevertheless,
to the author's knowledge, the effect of internal dissipation processes in turbulent flows has not
been studied yet. Thus, the dumbbell model with internal viscosity provides a simple setting for
a qualitative understanding of this phenomenon.

The study consists of Brownian dynamics simulations in three-dimensional homogeneous and isotropic turbulence and focuses on the 
statistics of polymer extension and the coil--stretch transition. The numerical results are explained
by adapting the theory in Ref.~\citenum{bfl00} to a dumbbell with internal viscosity.
In addition,
a fully analytical solution for a stochastic velocity gradient supports the interpretation of the results.
Finally, the concluding section discusses the experimental evidence for the effect of internal friction on single-polymer dynamics 
and identifies a phenomenon, namely the steepness of the probablity distribution
of the end-to-end distances, that can be attributed to
internal friction and not to other forces usually included in bead-spring chains, such as hydrodynamic and excluded-volume interactions
or a conformation-dependent drag.

\section{Model and methods}

\subsection{Elastic dumbbell with internal viscosity}

The polymer is described as an elastic dumbbell.
The extension and orientation are specified by the vector $\bm q$ that connects the two beads
and represents the end-to-end separation vector of the polymer.
Internal viscosity is introduced in the dumbbell
model by adding the resistive force 
$\bm F_{\rm iv}=-\phi (\bm q\cdot\overset{\bm .}{\bm q})\bm q/q^2$ to the equation for $\bm{q}$
(see Refs.~\citenum{kuhn,l88}).
Here $q=\vert\bm q\vert$ and
$\phi$ is termed the internal viscosity coefficient. The resistive force is 
parallel to $-\bm{q}$ and has a magnitude proportional to $dq/dt$.
Thus, for a finitely extensible
nonlinear elastic (FENE) dumbbell with internal viscosity the evolution equation for the connector vector is\cite{s92,hs95} 
\begin{equation}
\label{eq:dumbbell}
\dot{q}_i = A_i + C_{ijk}\, \kappa_{jk}(t) + B_{ij}\, \dot{W}_j(t)
\end{equation}
with
\begin{subequations}
\begin{eqnarray}
A_i &=& -\frac{q_i}{2\tau(1+\epsilon)(1-q^2/L^2)}-\frac{\epsilon}{1+\epsilon}\frac{4KT}{\zeta}\frac{q_i}{q^2},
\\[2mm]
B_{ij} &=& \sqrt{\frac{4KT}{\zeta}}\left[\delta_{ij} - \left(1-\sqrt{\frac{1}{1+\epsilon}}\right)\frac{q_iq_j}{q^2}\right],
\\[2mm]
C_{ikl} &=& \left(\delta_{ik}-\frac{\epsilon}{1+\epsilon}\frac{q_iq_k}{q^2}\right)q_l,
\end{eqnarray}%
\label{eq:coeff-dumbbell}%
\end{subequations}%
where $i,j,k=1,2,3$ and summation over repeated indices is understood,
$\tau$ is the relaxation time of the polymer, $L$ is its contour length,
$\zeta$ is the drag coefficient of the beads, $K$ is the Boltzmann constant, $T$ is temperature, 
$\kappa_{ij}=\partial u_i/\partial x_j$ is the velocity gradient at the position of the center of mass,
and $\bm{W}(t)$ is the three-dimensional Brownian motion [in Eq.~\eqref{eq:dumbbell} 
the noise term is interpreted in the It\^o sense\cite{hs95}]. The equilibrium length of the dumbbell, 
defined as the standard deviation of $q$ for $\bm\kappa=0$, is $q_{\rm eq}=\sqrt{12KT\tau/\zeta}$.

The parameter $\epsilon=2\phi/\zeta$ describes the ratio of internal viscosity to the hydrodynamic drag.
The usual FENE dumbbell model is recovered for $\epsilon=0$, whereas for infinite $\epsilon$
Eq.~\eqref{eq:dumbbell} yields the rigid dumbbell model.\cite{l88} In the literature, $\epsilon$ is typically
taken between 0 and 10.

The balance between polymer stretching and relaxation is
measured by the Weissenberg number, which in a chaotic flow 
is commonly defined as $\mathrm{Wi}=\lambda\tau$, where $\lambda$ is the maximum Lyapunov exponent of the flow, {\it i.e.} the average 
exponential rate at which fluid particles separate.

Here Eq.~\eqref{eq:dumbbell} is studied under the assumption that $\bm\kappa(t)$ is the gradient of a turbulent velocity field.
It is worth mentioning that even though
Eq.~\eqref{eq:dumbbell} assumes a linear velocity field, it remains appropriate for turbulent flows, because
the length of a polymer is generally shorter than the viscous dissipation scale, which is
the smallest length scale in such flows.

\subsection{Brownian dynamics simulations}
\label{sect:BD}

The effect of internal viscosity is studied by using
a database of Lagrangian trajectories in homogeneous and isotropic, incompressible turbulence generated at ICTS, Bangalore\cite{jr17}. 
Although an isotropic turbulent flow has zero mean strain rate, line elements are stretched exponentially with an asymptotic rate
$\lambda$. Thus, a polymer in an isotropic turbulent flow experiences strong stretching events that can
unravel it completely.\cite{bfl00,jc07,wg10}
The database was obtained
by tracking the positions of $10^5$ fluid particles and calculating $\bm\kappa(t)$ along their trajectories 
in a direct numerical simulation of the three-dimensional Navier--Stokes equations over a periodic cube and at Taylor-microscale Reynolds 
number ${\rm R}_\lambda=111$
(see Ref.~\citenum{jr17} for more details). The time series of $\bm\kappa(t)$ is then inserted into Eq.~\eqref{eq:dumbbell}.
This procedure assumes that the centre of mass of a polymer moves along a fluid trajectory; therefore the effect of thermal
noise on the motion of the centre of mass is disregarded. Such an assumption 
is justified in a turbulent flow, because thermal diffusion is 
negligible compared to turbulent diffusion.

\begin{figure*}[t]
\setlength{\unitlength}{\textwidth}
\includegraphics[width=0.325\textwidth]{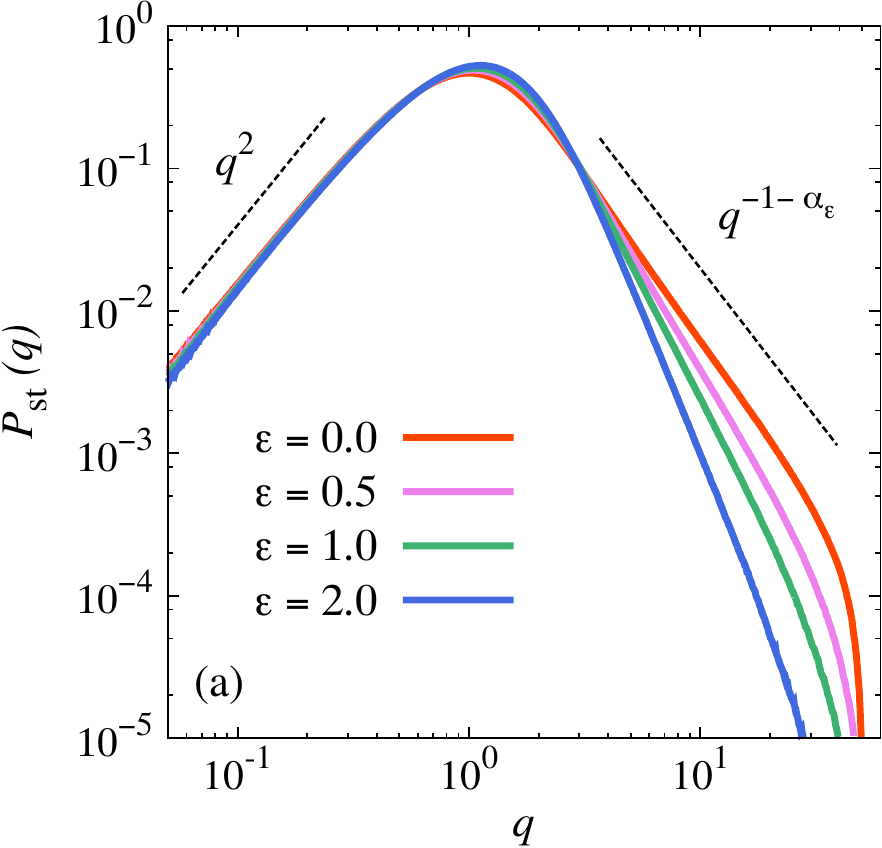}\hfill%
\includegraphics[width=0.325\textwidth]{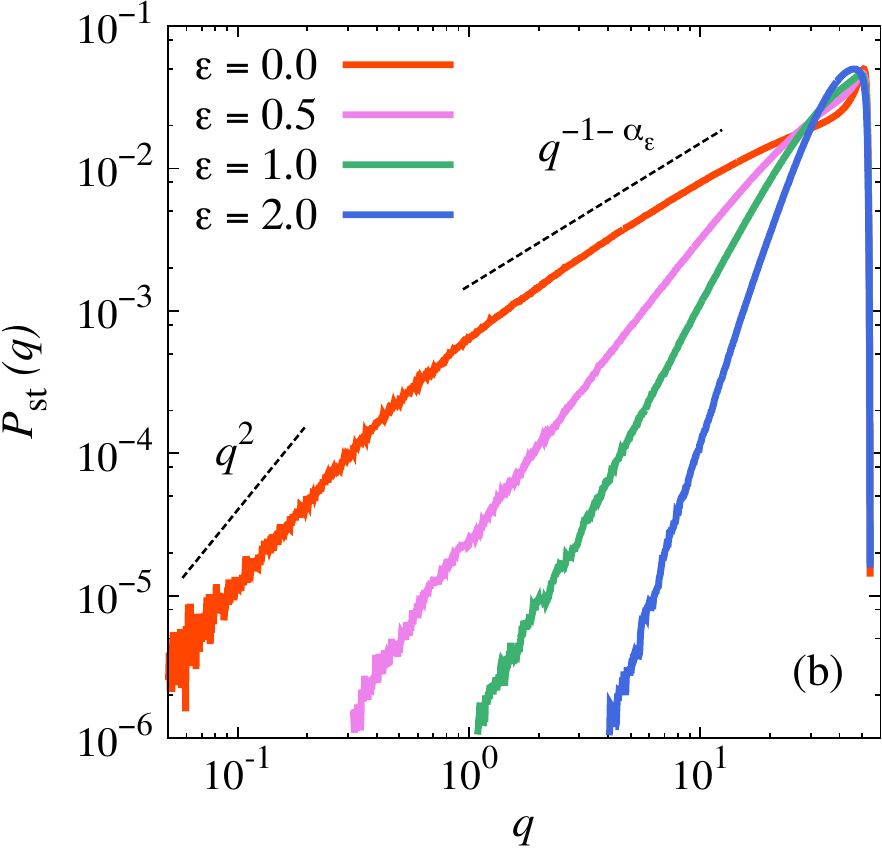}\hfill%
\includegraphics[width=0.325\textwidth]{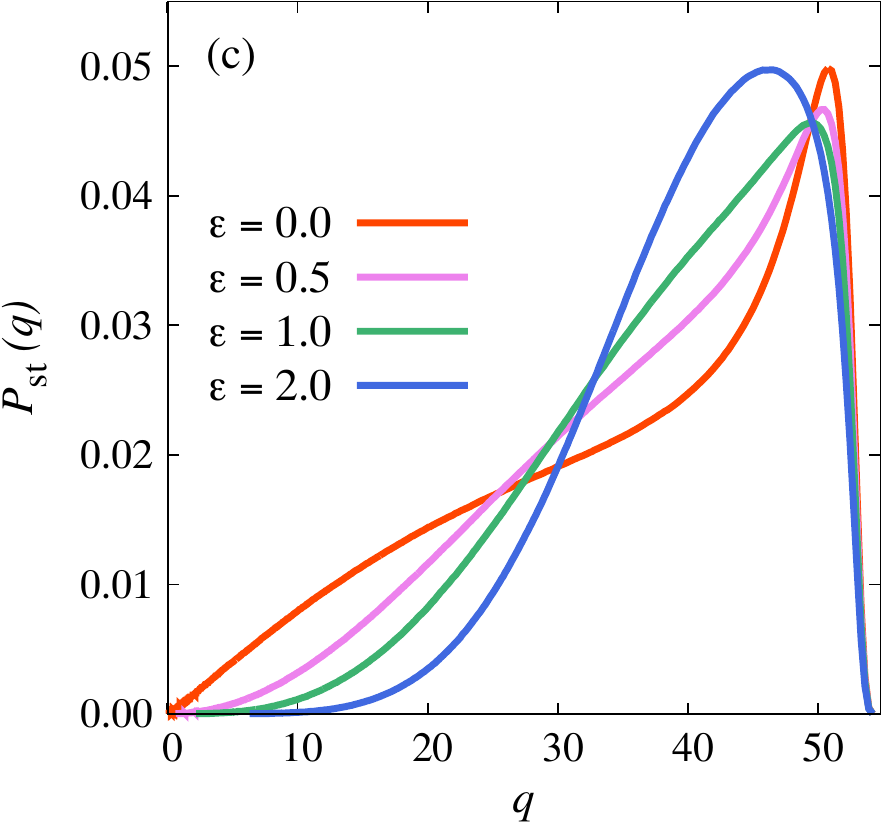}%
\caption{Stationary PDF of $q$ for different values of $\epsilon$ and (a) $\Wi=0.35$ and (b) $\Wi=2$. Panel (c) is the same as panel (b) but on a  
linear scale.}
\label{fig:1}
\end{figure*}

For the numerical integration of Eq.~\eqref{eq:dumbbell},
Ref.~\citenum{kcp18} proposes a semi-implicit predictor--corrector scheme, which
is adapted from an analogous algorithm initially derived for $\epsilon=0$.\cite{o96}
However, in the present setting
the Euler--Maruyama method supplemented with a rejection algorithm\cite{o96} proved accurate 
enough to prevent extensions greater than $L$ (for $\epsilon=0$ and the largest value of Wi---the least
favourable case---only 0.02\% of the time steps were rejected).
In the Lagrangian database, the velocity gradient was saved at a time interval $\Delta t=4\times 10^{-3}$. The time step used for the integration
of Eq.~\eqref{eq:dumbbell} is $dt=4\times 10^{-4}$;  a linear interpolation 
between two subsequent values of the velocity gradient is therefore required.
In Sect.~\ref{sect:simulations}, the contour length is $L=\sqrt{3\times 10^3}$ and $K$, $T$, $\zeta$, are such that 
$q_{\rm eq}$ is unity. The ratio $L^2/q_{\rm eq}^2$ is thus
comparable to that of long DNA molecules\cite{sbsc03} and is the same as that used in Refs.~\citenum{jc07,wg10}.
In Sect.~\ref{sect:alpha}, $L$ is taken unrealistically large, namely $L=10^3$, in order accurately to
resolve the power-law behaviour of the distribution of $q$.
The Weissenberg number is varied between 0.05 and 8, while $\epsilon$ is taken between 0 and 2.

\section{Results and discussion}

\subsection{Polymer stretching in isotropic turbulence: the effect of internal viscosity}
\label{sect:simulations}

The statistics of polymer stretching is described in terms of 
the stationary probability density function (PDF) of the polymer end-to-end distance, here denoted as $\pst(q)$.
It was already mentioned in Sect.~\ref{sect:introduction} that if $\epsilon=0$, analytical,\cite{bfl00} 
experimental,\cite{gcs05,ls10,ls14} and
numerical\cite{eks02,bcm03,pt05,wg10,bmpb12,gpp15}
studies of the dumbbell model have shown
that $\pst(q)$ behaves as a power of $q$ for $q_{\rm eq}\ll q\ll L$, with a slope that is negative for small Wi, motonically
increases as a function of Wi, and crosses $-1$ when the Weissenberg number takes the critical value $\Wicr=1/2$
(note that the present definition of $\tau$, and hence of Wi, differs by a factor of 2 from that used in some of the references
cited above).
Thus, in the $L\to\infty$ limit (linear polymer), $\pst(q)$ is no longer normalisable for $\Wi\geqslant \Wicr$, 
and  this is interpreted as the indication
of the coil--stretch transition occurring at $\Wi=\Wicr$.
The power-law behaviour of $\pst(q)$ means that the PDF is not dominated by a peak about its mean,
but the distribution of polymer extensions is broad.
The transition is characterized by a rapid increase of the mean extension and a sharp maximum in the 
coefficient of variation of $q$, defined as $\sigma/\langle q\rangle$, where $\sigma$ is the standard deviation of $q$
 (see Refs.~\citenum{gcs05,wg10,mav05}). The latter behaviour is a further indication of the 
breadth of the distribution and the heterogeineity of polymer configurations in a turbulent flow.

\begin{figure*}[!t]
\setlength{\unitlength}{\textwidth}
\includegraphics[width=0.511\textwidth]{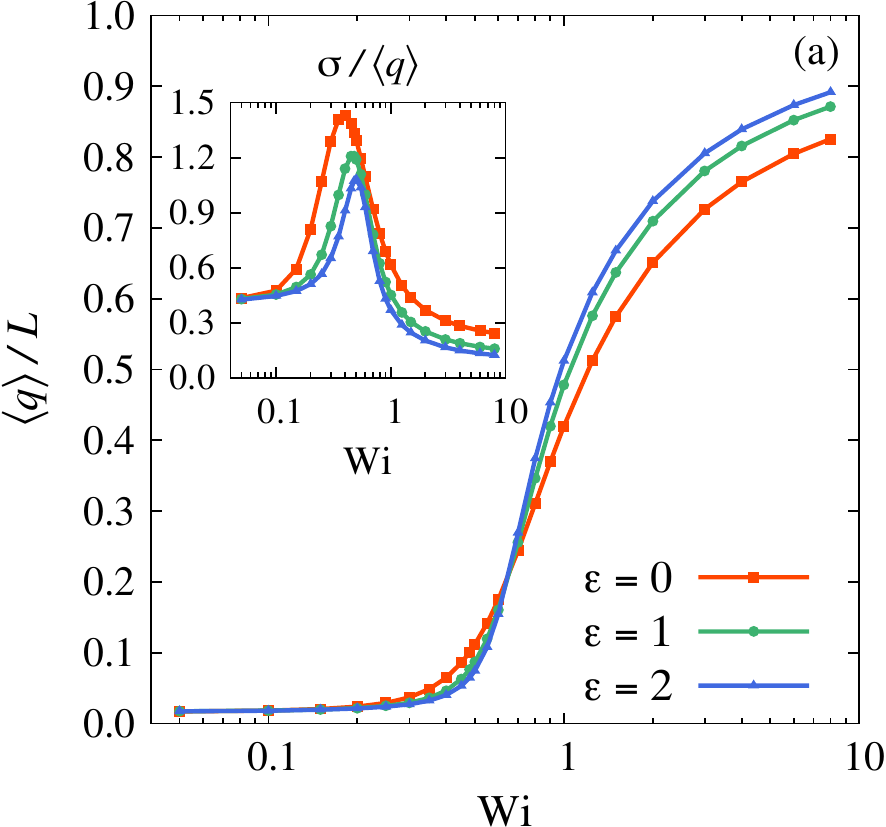}\hfill%
\includegraphics[width=0.489\textwidth]{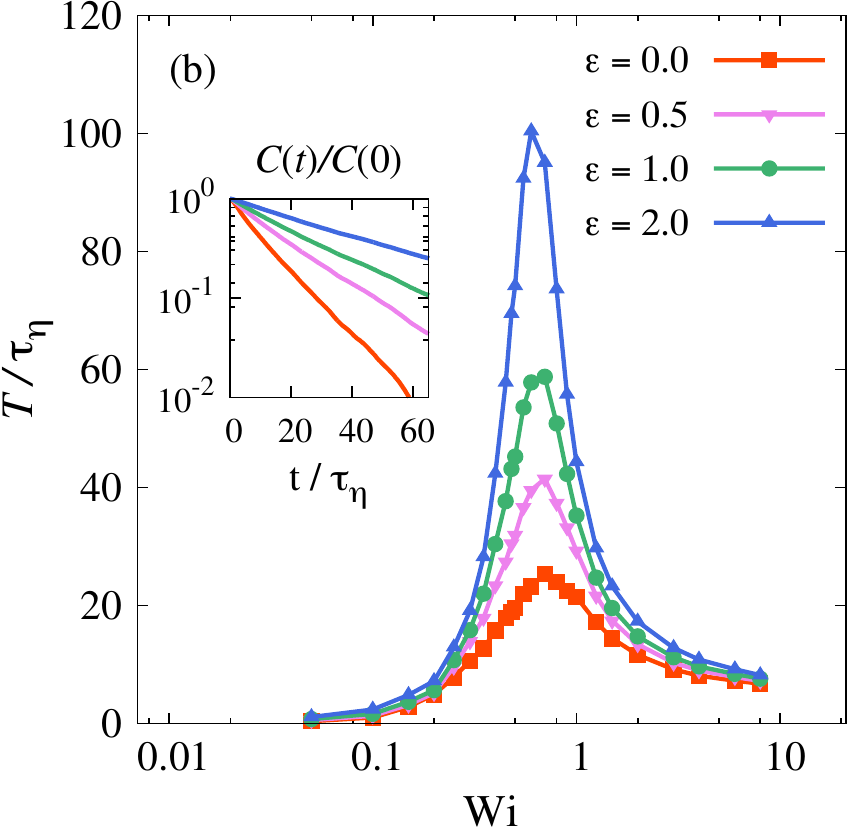}
\caption{(a) Mean extension rescaled by the contour length as a function of Wi for different values of $\epsilon$. The inset shows the coefficient
of variation of $q$ for the same values of Wi and $\epsilon$. (b) Correlation time of the polymer end-to-end distance  as a function of Wi for different values of $\epsilon$. The inset shows the autocorrelation of the polymer extension vs time for $\Wi=0.6$.
The time scale $\tau_\eta$ is the Kolmogorov time.}
\label{fig:mean}
\end{figure*}

Figure~\ref{fig:1} shows that, in the presence of internal viscosity, $\pst(q)$ continues to behave as a power of $q$ for intermediate extensions, but
the slope of the power law changes significanly
with $\epsilon$. Internal viscosity indeed makes the power steeper: the PDF falls faster than for $\epsilon=0$ at small Wi and rises faster at large Wi.
As a consequence, the mean polymer extension displays a sharper transition from the coiled to the stretched state as $\epsilon$ is increased, and its asymptotic value is 
larger at higher $\epsilon$ (see Fig~\ref{fig:mean}(a)). 
At the same time, the dispersion of the PDF around the mean is reduced by internal viscosity,
as is quantified by a systematically smaller coefficient of variation for $\epsilon>0$ (inset of Fig.~\ref{fig:mean}(a)).

This behaviour of $\pst(q)$ may be inferred from the fact that, in a turbulent flow, large deviations from the mean extension are the cumulative result of 
strong fluctuations of the velocity
gradient, and the effect of internal viscosity is to attenuate the response of the polymer to sudden variations in the velocity gradient.
However, a rigorous explanation of this phenomenon will be given in the following sections.

The coil--stretch transition also manifests itself in a significant
increase of the correlation time of polymer the end-to-end distance.\cite{wg10} 
A related phenomenon is the slowing down of the equilibration dynamics of the polymer in the flow.\cite{cpv06}
%this behaviour was explained by examining the effective potential $E(q) \propto -KT\ln\pst(q)$ that determines the stationary PDF of $q$.
%For $\Wi$ close to $\Wicr$, the potential displays a very wide well, which indicates a large variety of accessible configurations
%near the coil--stretch transition and hence a longer time needed for the system to explore the entire phase space.
If $C(t)=\langle q(t)q(0)\rangle -\langle q(t)\rangle^2$ is the autocorrelation function of the end-to-end distance, the
correlation time is defined as $T=\int_0^\infty dt\,C(t)/C(0)$.
The inset of 
Fig.~\ref{fig:mean}(b) shows that $C(t)$ decays approximately as an exponential function, as was already observed for $\epsilon=0$
(see Ref.~\citenum{wg10}). However,
internal viscosity strongly amplifies the aforementioned increase of the correlation time near the coil--stretch transition:
$T$ displays
a higher and higher peak near $\Wicr$ as $\epsilon$ grows (in Fig.~\ref{fig:mean}(b), time is rescaled by
the Kolmogorov time $\tau_\eta$, which is the time scale associated with viscous dissipation in turbulent flows).

Finally, the orientation dynamics of polymers in isotropic turbulence  has also attracted some attention.\cite{wg10,bmpb12,vpbt15}
%For sufficiently large Wi, the dynamics coincide with that
%of a rod,\cite{pw11} \textit{i.e.} $\bm q$ displays a moderate alignment with the middle eigenvector of the velocity gradient\cite{wg10}
%and a strong alignment with the vorticity\cite{vpbt15} $\bm\omega=\nabla\times\bm u$.
%If thermal noise is disregarded, the evolution of the orientation indeed decouples from that of the extension
%and coincides with that of a rigid rod~\cite{orl82,bfl00} (see also Eqs.~\eqref{eq:decoupling} below).
Internal viscosity obviously does not affect the orientation of a polymer directly, because $\bm F_{\rm iv}$ is parallel to $\bm q$.
However, it may in principle do so indirectly, since it modifies the statistics of $q$.%, and for small extensions
%the aforementioned decoupling, and hence the analogy with a rod, do not hold.
The numerical results (not shown) indicate that internal viscosity causes a mild reduction of the alignment of the polymer with 
the vorticity
only for Wi smaller than $\Wicr$ and close to $\Wicr$. The effect on the orientation dynamics is otherwise negligible.

\subsection{An exactly solvable model}
\label{sect:BK}

The stationary PDF of
$q$ can be calculated exactly if the turbulent velocity gradient is modelled as a
a stochastic tensor with suitable statistical properties.
In the Batchelor regime of the three-dimensional Kraichnan model,\cite{fgv01} 
$\bm\kappa(t)$ is an isotropic traceless tensorial white noise, 
which means that  $\boldsymbol{\kappa}(t)$ is Gaussian, has zero mean, and two-time correlation 
\begin{equation}
\langle\kappa_{ij}(t)\kappa_{kl}(t')\rangle=\mathcal{K}_{ijkl}\delta(t-t'),\qquad
i,j,k,l=1,2,3
\end{equation} 
with
$\mathcal{K}_{ijkl}= \lambda(4\delta_{ik}\delta_{jl}-\delta_{ij}\delta_{kl}-\delta_{il}\delta_{jk})/3$.
This stochastic model of the velocity gradient has been widely used in the study of
turbulent transport~\cite{fgv01} and was applied for the first time to single-polymer dynamics in Ref.~\citenum{c00}.
With this choice of $\boldsymbol{\kappa}(t)$, the velocity gradient plays the role of
a multiplicative noise in the second term on the right hand side
of Eq.~\eqref{eq:dumbbell} and is interpreted in the Stratonovich sense.\cite{c00,fgv01}

By using the methods presented in Ref.~\citenum{pav16}, it can be shown that if $\bm\kappa(t)$ is as above,
then the PDF of the vector $\bm{q}$, denoted as $f(\bm q,t)$,
satisfies the Fokker--Planck equation
\begin{equation}
\label{eq:FPE-q}
\frac{\partial f}{\partial t} =
-\frac{\partial}{\partial q_i}\left(V_i f\right)
+\frac{1}{2}\frac{\partial^2}{\partial q_i\partial q_j}\left(D_{ij}f\right)
\end{equation}
with drift and diffusion coefficients
\begin{equation}
V_i =  A_i +\frac{1}{2}\mathcal{K}_{klmn}C_{jmn} \frac{\partial C_{ikl}}{\partial q_j},\quad
D_{ij} = B_{ik}B_{jk}+\mathcal{K}_{klmn}C_{ikl}C_{jmn},
\end{equation}
where $A_i$, $B_{ij}$, $C_{ijk}$ have been defined in Eqs.~\eqref{eq:coeff-dumbbell}.
Summation over repeated indices is assumed also in this section.
The change to spherical coordinates $\bm{q}=(q\sin\theta\cos\varphi,q\sin\theta\sin\varphi,q\cos\theta)$ transforms
Eq.~\eqref{eq:FPE-q} into a Fokker--Planck equation for $f(q,\theta,\varphi,t)$ with coefficients:
\begin{eqnarray*}
V_q &=& \frac{4KT}{(1+\epsilon)\zeta q}-\frac{q}{2(1+\epsilon)\tau(1-q^2/L^2)}+\frac{(4+3\epsilon)\lambda q}{3(1+\epsilon)^2},
\\[1mm]
V_\theta &=& 2\left(\frac{KT}{\zeta q^2}+\frac{\lambda}{3}\right)\cot\theta,
\quad
D_{\theta\theta} = \frac{4KT}{\zeta q^2} + \frac{4\lambda}{3},
\\[1mm]
D_{qq} &=& \frac{4KT}{(1+\epsilon)\zeta} + \frac{2\lambda q^2}{3(1+\epsilon)^2}, \quad
D_{\varphi\varphi} = 4\left(\frac{KT}{\zeta q^2}+\frac{\lambda}{3}\right)\csc^2\theta,
\end{eqnarray*}
and $V_\varphi=D_{q\theta}=D_{q\varphi}=D_{\theta\varphi}=0$ (see Ref.~\citenum{r89}, p.~88,
for the transformation rules of a Fokker--Planck equation under a change of variables). 
Taking into account the statistical isotropy of the flow,
it is now assumed that the stationary PDF of $\bm{q}$ is of the form
$f_{\rm st}(q,\theta,\phi)=\pst(q)\sin\theta$. By replacing this expression into the Fokker--Planck
equation for $f(q,\theta,\phi,t)$, it is then found that $\pst(q)$ satisfies the equation
\begin{equation}
\label{eq:FPE-1D}
2\partial_q(V_q \pst)=\partial^2_q (D_{q}\pst),
\end{equation}
which is solved with a reflecting boundary condition in $q=0$. This implies that in the steady state
the probability current vanishes everywhere.\cite{r89}
The solution of Eq.~\eqref{eq:FPE-1D}
corresponding to zero probability current is\cite{r89} $\pst(q) \propto D_{qq}^{-1} \exp\left(2 \int V_q/D_{qq}\,dq\right)$, whence the analytical expression of $\pst(q)$ is
\begin{equation}
\pst(q) \propto q^2 \left[1+\frac{2\Wi}{(1+\epsilon)}\,\frac{q^2}{q_{\rm eq}^2}\right]^{\frac{3}{2}\epsilon-\gamma_{\epsilon}}\left(1-\frac{q^2}{L^2}\right)^{\gamma_\epsilon}
\label{eq:pdf-exact}
\end{equation}
with 
\begin{equation}
\gamma_\epsilon^{-1} = \frac{2}{3}\left[\frac{q_{\rm eq}^2}{L^2} + \frac{2\Wi}{(1+\epsilon)}\right].
\end{equation}
If $\epsilon$ is set to zero, the above PDF reduces to that found in Ref.~\citenum{mav05} for a polymer with zero internal viscosity
in the Batchelor--Kraichnan flow.
At small $q$, the PDF is proportional to $q^2$, because the dynamics is dominated by thermal fluctuations. At very large $q$, the last term in Eq.~\eqref{eq:pdf-exact},
which originates from the nonlinear elastic force, introduces a cut-off at the length $L$.
For $q_{\rm eq}\ll q\ll L$, the stationary PDF of $q$ behaves as $\pst(q)\sim q^{-1-\alpha_\epsilon}$ with
\begin{equation}
\label{eq:alpha-example}
\alpha_\epsilon = \frac{3}{2}\,(1+\epsilon)\left(\frac{1}{\Wi}-2\right).
\end{equation}
The factor $1+\epsilon$ has the effect of reducing the slope of $\pst(q)$ for $\Wi<\Wicr$ and increasing it for $\Wi>\Wicr$.
This makes $\pst(q)$ narrower and the coil--stretch transition sharper, although it does
not modify $\Wicr$, which is defined as the value of Wi at which $\alpha_\epsilon$ vanishes. Thus,
the stochastic model captures the effect of internal viscosity on the steady-state statistics of polymer extension
as observed in the Brownian dynamics simulations and provides an analytical tool for the study of internal viscosity
in turbulent flows.

\subsection{Predictions for a general random flow}
\label{sect:alpha}

The behaviour of $\pst(q)$ observed in the Brownian dynamics simulations and reproduced by the stochastic model can be predicted
for a general random flow by invoking the theory in Ref.~\citenum{bfl00}. This is
briefly recalled below in the version provided in Ref.~\citenum{bcm03}, which uses the generalized Lyapunov exponents. 

Let $\bm\ell(t)$ be a line element in a random flow. Its time evolution is given by the equation $\dot{\bm\ell}(t) = \bm\kappa(t)\cdot\bm\ell(t)$, which
in turn yields the following equation for the length of the line element:
\begin{equation}
\label{eq:line}
\dfrac{d}{dt}\ln\ell=\widehat{\bm{\ell}}\cdot\bm{\kappa}(t)\cdot\widehat{\bm{\ell}}
\end{equation}
with $\widehat{\bm{\ell}}=\bm{\ell}/\ell$.
The $p$-th generalized Lyapunov exponent is defined as\cite{cpv93,cc2010} 
\begin{equation}
\label{eq:Lp}
\mathcal{L}(p) = \lim_{t\to\infty}\frac{1}{t}\ln\left\langle\left[\frac{\ell(t)}{\ell(0)}\right]^p\right\rangle,
\end{equation}
where $\langle\cdot\rangle$ denotes the average over the statistics of the velocity field.
$\mathcal{L}(p)$ represents the rate of exponential growth of the $p$-th moment of $\ell(t)$. 
It is a positive and convex function of $p$ and satisfies $\mathcal{L}(0)=\mathcal{L}(-d)=0$, where $d$ is the space dimension. 
In addition, $\mathcal{L}'(0)=\lambda$.

References~\citenum{bfl00} and~\citenum{bcm03} express $\pst(q)$ in terms of $\mathcal{L}(p)$ (or its Legendre transformation). 
It is first observed that if $\epsilon=0$ and thermal noise is disregarded, the end-to-end distance and the orientation of a {\it linear} polymer
evolve according to the following equations:\cite{orl82,bfl00} 
\begin{subequations}
\label{eq:decoupling}
\begin{eqnarray}
\label{eq:log-0}
\dfrac{d}{dt}\ln q &=& \beta(t)-\frac{1}{2\tau} \qquad\qquad (\epsilon=0)\\
\dfrac{d\widehat{\bm q}}{dt} &=& \bm{\kappa}(t)\cdot\widehat{\bm{q}}-\beta(t)\widehat{\bm q}
\label{eq:qhat-0}
\end{eqnarray}
\end{subequations}
with $\beta(t)=\widehat{\bm{q}}\cdot\bm{\kappa}(t)\cdot\widehat{\bm{q}}$.
The similarity between Eq.~\eqref{eq:log-0} and Eq.~\eqref{eq:line} makes it clear that the 
statistics of $q$ must be related to the generalized Lyapunov exponents of the flow.
Extensions much greater than $q_{\rm eq}$ are observed after the polymer has experienced large values of $\beta(t)$.
Thus, $q$ is expressed in terms of $\beta(t)$ by writing the first of Eqs.~\eqref{eq:log-0} in integral form,
and then the probability of large values of $\beta(t)$ is approximated with its large-deviations form to find:
\begin{equation}
\label{eq:alpha-0}
\pst(q)\sim q^{-1-\alpha_0} \quad \text{with} \quad \alpha_0=2\tau\mathcal{L}(\alpha_0)
\end{equation}
for $q_{\rm eq}\ll q\ll L$.
The value of $\alpha_0$ is therefore sought as the nonzero intersection of the straight line $\alpha_0/2\Wi$ with
the function $\mathcal{L}(\alpha_0)/\lambda$.
By using the aforementioned properties of $\mathcal{L}(p)$ as a function of $p$, it is easy to see that 
$\alpha_0$ is positive for small Wi and decreseas with Wi, until it vanishes for $\Wi=\Wicr$. It then becomes negative for $\Wi>\Wicr$.
Close to $p=0$, the generalized Lyapunov exponent can be expanded as
$\mathcal{L}(p)=\lambda p+\varDelta p^2/2+O(p^3)$ with
$\varDelta = \int \left(\left\langle\beta(t)\beta(t')\right\rangle - \lambda^2\right)dt'$.
This expansion allows the explicit calculation of $\alpha_0$ for Wi near to $\Wicr$:
\begin{equation}
\label{eq:alpha-kraichnan}
\alpha_0=\frac{\lambda}{\varDelta}\left(\frac{1}{\Wi}-2\right).
\end{equation}
In particular, the latter expression shows that $\Wicr=1/2$.
Finally, the limit of $\alpha_0$ for infinite Wi is obtained when the straight line $\alpha_0/2\Wi$ is parallel to the
horizontal axis, whence $\lim_{\mathrm{Wi}\to\infty} \alpha_0=-d$.

It is worth mentioning that Eq.~\eqref{eq:alpha-0} holds under very mild assumptions on the random flow, namely that
the correlation time of $\beta(t)$ is finite.\cite{bfl00} Moreover, even though Eq.~\eqref{eq:alpha-0} is derived for a
dumbbell, Ref.~\citenum{wg10} has shown that the steady-state statistics of the end-to-end distance is the same for a dumbbell and
a chain with multiple beads, provided that 
a suitable mapping between the parameters of the two systems is applied. 
Hence the validity of Eq.~\eqref{eq:alpha-0} is not restriced to the dumbbell model.

It is now discussed how internal viscosity modifies the above predictions.
%(the definition of $\tau$ in Ref.~\citenum{bfl00} differs from ours by a factor of 2).
If $\epsilon>0$, the analogue of Eq.~\eqref{eq:log-0} can be obtained by multiplying Eq.~\eqref{eq:dumbbell} by $q_i$, 
neglecting the noise term, summing over $i$, and dividing by $q^2$ to find:
\begin{equation}
\label{eq:log}
\dfrac{d}{dt}\ln q=\frac{1}{1+\epsilon}\left[\beta(t)-\frac{1}{2\tau}\right] \qquad (\epsilon\geqslant 0).
\end{equation}
Constrastingly, Eq.~\eqref{eq:qhat-0} is unchanged.
Therefore, for $\epsilon> 0$,
the time evolution of $q(t)$ is the same as that of a polymer with $\epsilon=0$, provided that
$\tau$ is multiplied by $1+\epsilon$ and
$\beta(t)$ is rescaled by the same quantity. It follows immediately that
$\pst(q)$ must display a power-law behaviour also in the presence of internal viscosity:
\begin{equation}
\pst(q)\sim q^{-1-\alpha_\epsilon} \quad (q_{\rm eq}\ll q\ll L)
\end{equation}
with an exponent that can be determined as follows. Rescaling $\beta(t)$ by $(1+\epsilon)$ is equivalent to considering the evolution of $q$ in a flow with
generalized Lyapunov exponents $\mathcal{L}_\epsilon(p)=\mathcal{L}(p/(1+\epsilon))$.
This can be seen by noting that the solution of Eq.~\eqref{eq:line} is $\ell(t)=\ell(0)\exp[
\int_0^t ds\:\widehat{\bm{\ell}}\cdot\bm{\kappa}(s)\cdot\widehat{\bm{\ell}}\,]$;
if this expression is
replaced into Eq.~\eqref{eq:Lp}, it follows that considering a flow with a rescaled $\beta(t)$ is the same as taking a moment of $\ell(t)$ of a rescaled order
in the original flow.
Hence, the equivalent of Eq.~\eqref{eq:alpha-0} for $\epsilon>0$ is
\begin{equation}
\frac{\alpha_\epsilon}{1+\epsilon}=2\tau\mathcal{L}_\epsilon(\alpha_\epsilon)=2\tau\mathcal{L}\left(\frac{\alpha_\epsilon}{1+\epsilon}\right).
\label{eq:alpha-epsilon}
\end{equation}
\begin{figure}[t]
\centering
\includegraphics[width=\columnwidth]{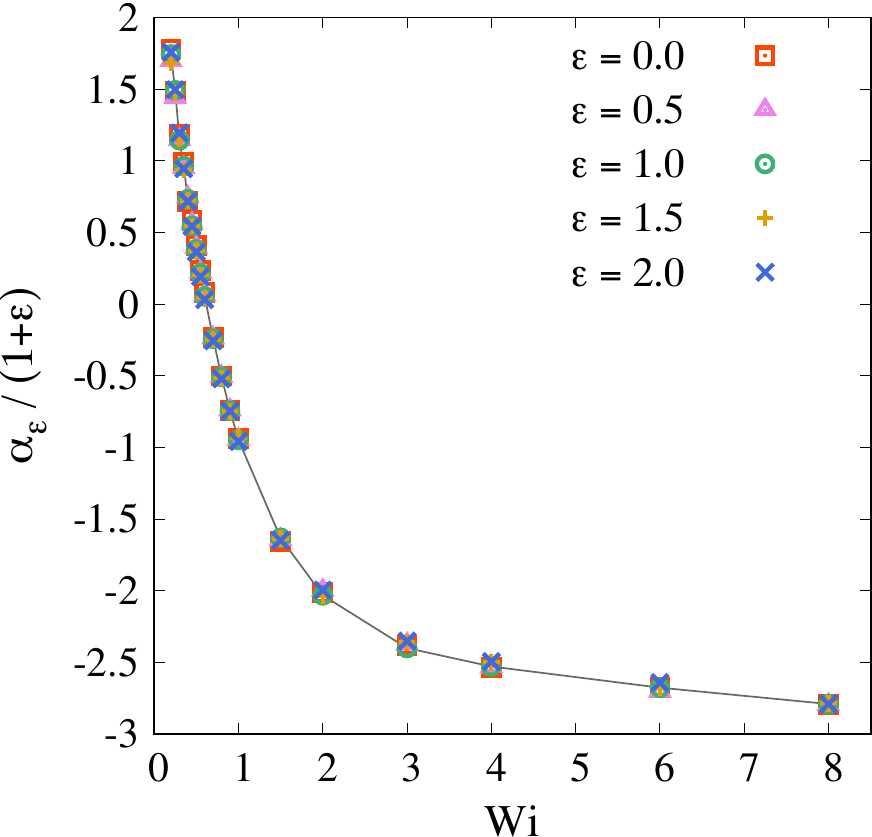}\hfill%
\caption{Exponent $\alpha_\epsilon$ rescaled by $1+\epsilon$ as a function of Wi for different values of $\epsilon$.}
\label{fig:3}
\end{figure}
Comparing Eq.~\eqref{eq:alpha-epsilon} with Eq.~\eqref{eq:alpha-0} finally yields 
\begin{equation}
\alpha_\epsilon=(1+\epsilon)\alpha_0.
\label{eq:alpha-epsilon-0}
\end{equation}
Therefore, the effect of internal viscosity on the PDF of polymer extensions
is to multiply $\alpha_0$ by a factor of $(1+\epsilon)$. Since the criterion for the coil--stretch transition in random flows is $\alpha_\epsilon=0$,
internal viscosity does not modify the critical Weissenberg number. However, the statistics of $q$ is affected. Indeed, $-1-\alpha_\epsilon < -1-\alpha_0$
when $\alpha_0<0$, \textit{i.e.} for $\Wi<\Wicr$, whereas $-1-\alpha_\epsilon > -1-\alpha_0$ in the opposite case. Thus, below the coil--stretch transition
the probability of large extensions is depleted by internal viscosity; above the transition it is the
small extensions that are disfavoured. As a result,
the mean extension is reduced when Wi is below $\Wicr$ and increased for $\Wi>\Wicr$, while the width of the PDF of the extension is systematically decreased by internal viscosity.
The coil--stretch transition therefore becomes sharper with increasing $\epsilon$.
Figure~\ref{fig:3} clearly illustrates the validity of Eq.~\eqref{eq:alpha-epsilon-0} by showing $\alpha_\epsilon$ rescaled
by $(1+\epsilon)$ from 
the Brownian dynamics simulations described in Sect.~\ref{sect:BD}
he value of $\alpha_\epsilon$ is estimated by fitting $\pst(q)$ for
$q_{\rm eq}\ll q\ll L$ with a power law (in order to obtain an accurate estimate, here the ratio of $L$ and $q_{\rm eq}$ is taken larger
than in Sect.~\ref{sect:simulations}, {\it i.e.} $L/q_{\rm eq}=10^3$).

In the Batchelor--Kraichnan flow studied in Sect.~\ref{sect:BK}, $\mathcal{L}(p)$ is exactly quadratic for all $p$ (see Ref.~\citenum{fgv01}). Hence the expression
for $\alpha_0$ given in Eq.~\eqref{eq:alpha-kraichnan} holds for all Wi and not only near the coil--stretch transition. 
In addition, $\lambda/\varDelta=d/2$ for this flow.~\cite{fgv01}
Therefore, Eq.~\eqref{eq:alpha-example} is an explicit example of the general relation  given
in Eq.~\eqref{eq:alpha-epsilon-0}.

\section{Summary and conclusions}

The effect of internal friction on polymer stretching in turbulent flows has been studied by considering an elastic dumbbell with a linear dashpot.
The results are based on Brownian dynamics simulations using a database of fluid trajectories in isotropic turbulence, an exact solution for a stochastic
velocity gradient, and a generalization of the large deviations approach of Ref.~\citenum{bfl00} that takes internal viscosity into account.
Although it does not modify the critical Weissenberg number for the coil--stretch transition, internal viscosity strongly affects the statistics of
polymer extension in two opposite ways below and above the transition. 
Its effect is indeed to multiply $\alpha_0$ by a factor of $(1+\epsilon)$.
This depletes 
the probability of large extensions below the transition
and the probability of small extensions above the transition, thus leading to a
sharpening of the transition itself.
Internal viscosity also enhances the peak of the correlation time of the extension near $\Wicr$, whereas it has a negligible effect on the orientation statistics
of the polymer.

It remains to consider the question of the experimental evidence for the phenomenon described here.
If internal viscosity is disregarded, the theory\cite{bfl00} predicts that, 
in the limit of very large Weissenberg numbers, $\alpha_0$ should tend to $-d$, and hence, in a three dimensional flow, 
$\pst(q)\sim q^2$ for $q_{\rm eq}\ll q\ll L$.
However, experiments~\cite{ls14} show PDFs as steep as $q^4$ when the Weissenberg number is large. The $\epsilon=0$ theory therefore does not
explain the shape of $\pst(q)$ in the large-Wi regime.
Contrastingly, if internal viscosity is taken into account, Eq.~\eqref{eq:alpha-epsilon-0} implies that
$\alpha_\epsilon\to-d(1+\epsilon)$ and hence $\pst(q)\sim q^{d(1+\epsilon)-1}$ as $\Wi\to\infty$.
Thus, internal friction provides a possible explanation for the steep behaviour
of $\pst(q)$ observed in experiments at large Wi. Moreover, the experimental slope $\pst(q)\sim q^4$ is
recovered by taking $\epsilon=0.67$, a value of $\epsilon$ which falls in the range typically
considered in studies of internal friction.

It is interesting to note that other forces that are usually included in bead-spring chain 
models cannot explain the steepness of $\pst(q)$ for large Wi.
Hydrodynamic interactions between the beads have the effect of delaying the unravelling of the polymer, and once
this is sufficiently stretched, they become negligible.
Therefore, in a turbulent flow, hydrodynamic interactions reduce the probability of large extensions for all Wi.
Excluded-volume interactions are short-range and do not impact the statistics of large polymer extensions.
A conformation-dependent drag, which interpolates bewteen the drag coefficient of a sphere in the coiled state and that of a thin cylinder in the
stretched state,\cite{h75} impacts the dynamics of the polymer around the coil--stretch transition, but has little 
effect at large Wi, when most of the polymers are highly stretched anyway.\cite{sbsc03,ssc04,cpv06}
In particular, by using the analytical results of Ref.~\citenum{cpv06}, it is easy to check that, in the Batchelor--Kraichnan flow,
$\pst(q)\sim q^2$ as $\Wi\to\infty$, even if the drag coefficient of the polymer depends on its conformation.
Thus, the present study identifies a phenomenon that is the unambiguous result of internal friction.

\section*{Acknowledgements}
The author is grateful to S.~S.~Ray for providing access to his database of turbulent Lagrangian trajectories
and to J.~R.~Picardo for useful suggestions.
The Brownian dynamics simulations  were  perfomed  at  Centre  de  Calculs  Interactifs  of  Universit\'e  C\^ote d’Azur.

%%%END OF MAIN TEXT%%%


\balance


%%%REFERENCES%%%
\bibliography{internal-viscosity} 
\bibliographystyle{rsc} %the RSC's .bst file

\end{document}